\documentclass{aastex}
\usepackage{emulateapj5}
\usepackage{epsfig}

\makeatletter
\newenvironment{inlinefigure}{%
\def\@captype{figure}%
\noindent\begin{minipage}{0.999\linewidth}\begin{center}}
{\end{center}\end{minipage}\smallskip}
\makeatother

\newcommand{\etal}{\mbox{et al.}}

\newcommand{\rxte}{{\it RXTE}}

\newcommand{\aqlxone}{Aql~X-1}

\begin{document}

\shortauthors{Muno \etal}
\shorttitle{Color-color Diagrams of LMXBs}

\title{How Do Z and Atoll X-ray Binaries Differ?}
\author{Michael P. Muno, Ronald A. Remillard, and Deepto 
Chakrabarty\altaffilmark{1}}
\affil{Department of Physics and Center for Space Research, 
       Massachusetts Institute of Technology, Cambridge, MA 02139}
\email{muno,rr,deepto@space.mit.edu}
\altaffiltext{1}{Alfred P. Sloan Research Fellow}

\begin{abstract}
Low-mass X-ray binaries containing weakly magnetized neutron stars may
 be divided into two classes, 
Z and atoll sources, based upon correlations between their X-ray timing
properties and the patterns which 
they trace in plots of two X-ray colors. In this paper, we examine 
color-color diagrams of eight atoll sources and four Z sources using data 
from the {\it Rossi X-ray Timing Explorer}. The five-year span of data we 
have examined is 
significantly longer than those of previous studies. We find that the previous 
clear distinction between color-color diagrams from atoll and Z sources
is an artifact of incomplete sampling, as
those atoll sources which are sampled over a wide dynamic range in intensity
($F_{\rm max}/F_{\rm min}\gtrsim 80$) trace three-branched color-color patterns
similar to the tracks for which Z sources are named. However, atoll sources
trace this pattern over a larger range of luminosity and on much longer time
scales than do Z sources, and exhibit much harder spectra when they are 
faint, which argues against any simple unification scheme for the two classes
of source.
\end{abstract}

\section{Introduction}
Low-mass X-ray binaries (LMXBs) containing weakly-magnetized neutron 
stars may be divided into two classes, Z and
atoll sources, based upon correlations between their spectral colors 
and Fourier timing properties at X-ray wavelengths \citep{hv89}. 
Plots of a ``hard'' color against a ``soft'' color from Z sources usually 
form a Z shape that 
is traced on time scales of hours to days (bottom panels of 
Figure~\ref{cconly}). Plots from atoll sources 
often resemble a band of points at constant hard color, with 
``islands'' of points appearing on time scales of 
weeks and months (see 4U~1820-303 in Figure~\ref{cconly}). Power spectra
from both types of sources may be described with similar broad-band 
noise components, but 
Z sources exhibit strong (up to 10\% rms) 
quasi-periodic oscillations (QPOs) between 1--60~Hz, while atoll 
sources do not (but see Wijnands \& van der Klis 1999; Psaltis, Belloni,
\& van der Klis 1999). The exact causes of the 
spectral and timing variability 
is still unknown, but it is thought that the differences between the 
two classes result from a higher rate of mass transfer in Z sources than 
atoll sources \citep[see][for a review]{vdk95}.

In this {\it Letter}, we examine the spectral 
changes that occur in Z and atoll sources, using the public archive of 
observations taken with the Proportional Counter Array \citep[PCA;][]{jah96} 
aboard the {\it Rossi X-ray Timing Explorer} (\rxte). 
This unprecedented database of PCA observations, which samples the spectral
variability of LMXBs on time scales 
from seconds to years, allows us to present a complete picture of the X-ray 
color changes that occur in these systems.

\section{Data Analysis}

We have obtained all of the publicly available \rxte\ PCA data as of 2001 
September (along with some proprietary data) for twelve neutron star 
LMXBs (Table~\ref{stats}).
Seven of these LMXBs have been previously classified as atoll sources 
and four as Z sources \citep{vdk95,rei00}, while one (GS~1826$-$238) is 
probably an atoll source since it prolifically produces 
thermonuclear X-ray bursts 
\citep{ube99,vdk95}. The public archive of pointed observations of 
these sources contains many dozens of observations 
between 3000 to 30,000 seconds long, spanning over five years.

We have examined data with 128 energy channels between 2--60~keV and 16 s 
time resolution from the PCA, and daily flux histories 
from the \rxte\ All-Sky Monitor \citep[ASM;][]{lev96}. We list some basic 
properties of the sources in Table~\ref{stats}. 
The variability 
is defined as the ratio of the maximum to minimum count rate observed with
the PCA between 2--18~keV, and is a lower limit to the true value. This 
sample of sources spans a factor of 100 in average luminosity, and a 
factor of 1000 in variability.

\begin{figure*}[t]
\centerline{\epsfig{file=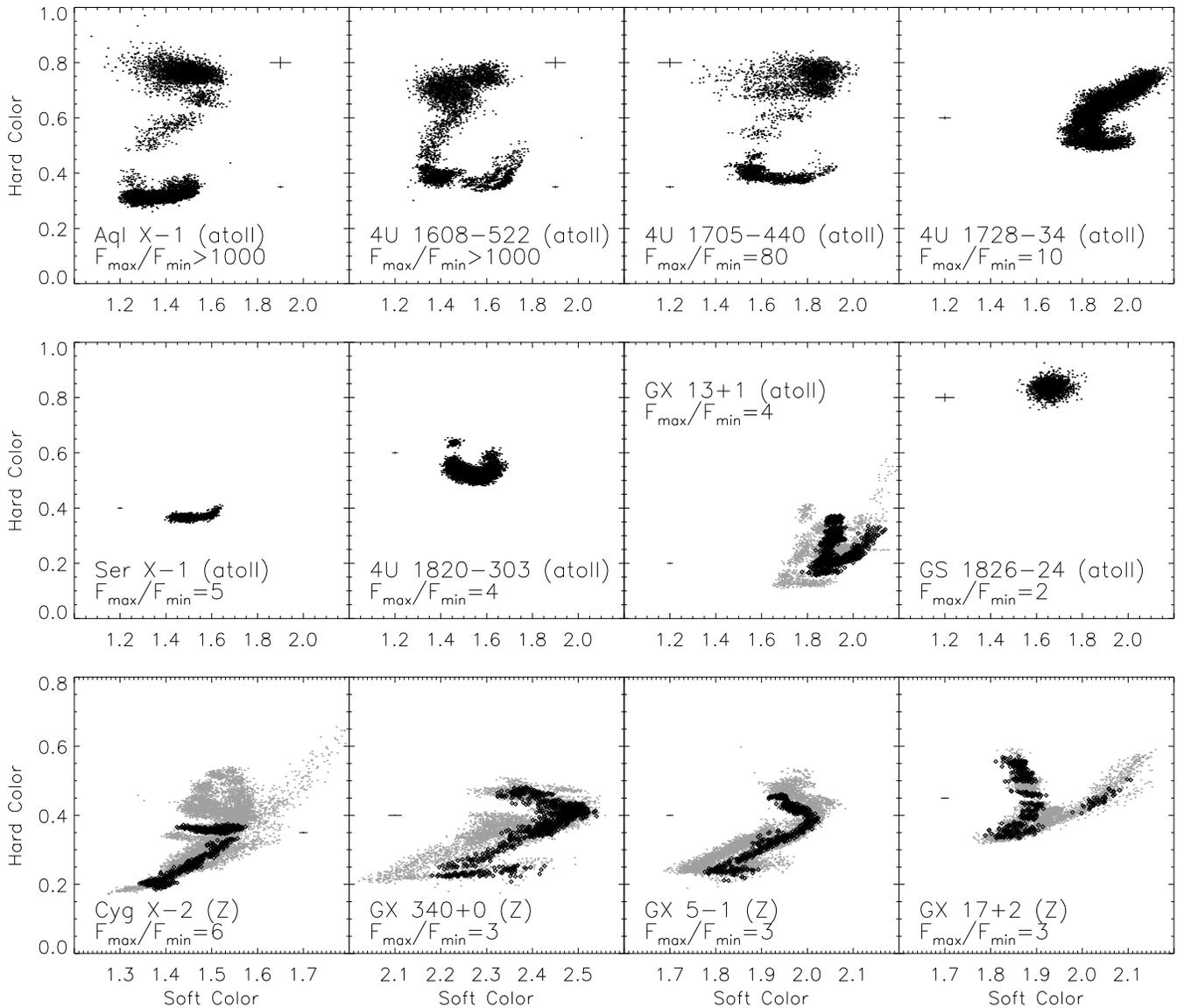,width=\linewidth}}
\caption{
Color-color diagrams from five years of pointed \rxte\ PCA observations. 
Typical uncertainties are indicated to the side of the data. The atoll  
sources that vary by the widest range in intensity trace Z-shaped tracks,
similar to those of Z sources. 
The remaining atoll sources trace portions of this complete track. The Z 
sources differ from the atoll sources in that 
they tend to be softer, and they trace their full range of spectral 
variability on time scales of days and with smaller intensity variations. 
GX~13$+$1 is unusual,
in that it has been previously classified as an atoll source, yet its
color-color diagram resembles that of the Z source GX~17$+$2 more than 
any other source. We have 
highlighted data spanning twenty days for sources that trace a complete
pattern in their X-ray colors on day-long time scales: 
MJD 50980--51000 for GX~13$+$1, MJD~50600--50620 in GX~340$+$0, 
MJD~50990--51010 for Cyg~X-2, and MJD 50530--50550 for GX~17+2. Each time
interval contains between 4--17 observations.}
\label{cconly}
\end{figure*}

We have defined hard and soft colors as the ratio of the background-subtracted 
detector counts in the (3.6--5.0)/(2.2--3.6) keV and the 
(8.6--18.0)/(5.0--8.6) keV energy bands, respectively, to produce color-color 
and color-intensity diagrams (Figures~\ref{cconly}--\ref{cctime}). 
We have used 64 s integrations to calculate the colors when the source 
intensity is above 100 counts s$^{-1}$, and 256 s integrations otherwise.
Gain changes in 
the five individual proportional counter 
units (PCUs) of the PCA over the course of the 
mission cause systematic variation in these count rates 
\citep[e.g.,][for a more detailed discussion]{hom01}. Therefore,
we have normalized the count rates from the Crab Nebula in each PCU to 
constant values for each energy band (totaling 2440 counts s$^{-1}$ 
PCU$^{-1}$ in the 2.2--18.0 keV band) using linear trends. 
When this correction is applied, the hard and soft colors from the Crab Nebula 
have values of 1.358 and 0.679, with standard deviations
of only 0.5\% and 0.1\% respectively. 

\section{Results}

We plot the color-color diagrams in Figure~\ref{cconly}.
The most striking aspect of Figure~\ref{cconly}
is that a Z-shape is formed in 
color-color diagrams from the
atoll sources that vary by the widest range in X-ray 
intensity: Aql~X-1, 4U~1608$-$522, 
and 4U~1705$-$44 (Table~\ref{stats}).
It has long been suspected that the complete color-color diagram from 
atoll sources forms a Z-shape \citep[e.g.][]{lan89}, but this is the first 
time 
that atoll sources have been adequately sampled over a wide enough range 
of luminosity to observe this pattern directly.
Several more sources that vary by 
no more than a factor of 10 in X-ray intensity appear to form portions of this 
pattern.  4U~1820$-$303 and 4U~1728$-$34 in Figure~\ref{cconly} 
exhibit only the bottom and diagonal 
portions of this pattern. Ser~X-1 exhibits only the bottom portion, and 
GS~1826$-$238 only the top (Figure~\ref{cconly}). 

The tracks from the Z sources Cyg~X-2, GX~17+2, GX~5$-$1, and GX~340$+$0 
(Figure~\ref{cconly}) are 
unmistakably Z-shaped, but are somewhat
different from those of the atoll sources.  In general, Z sources
are softer than atoll sources. Moreover, the traditional Z sources 
trace out a full track on time scales as short as a day, while 
atoll sources form their color-color diagrams on much longer time 
scales of 30-100 days. We have 
highlighted using darker symbols data spanning 20 days for the sources 
which trace their 

\begin{inlinefigure}
\centerline{\epsfig{file=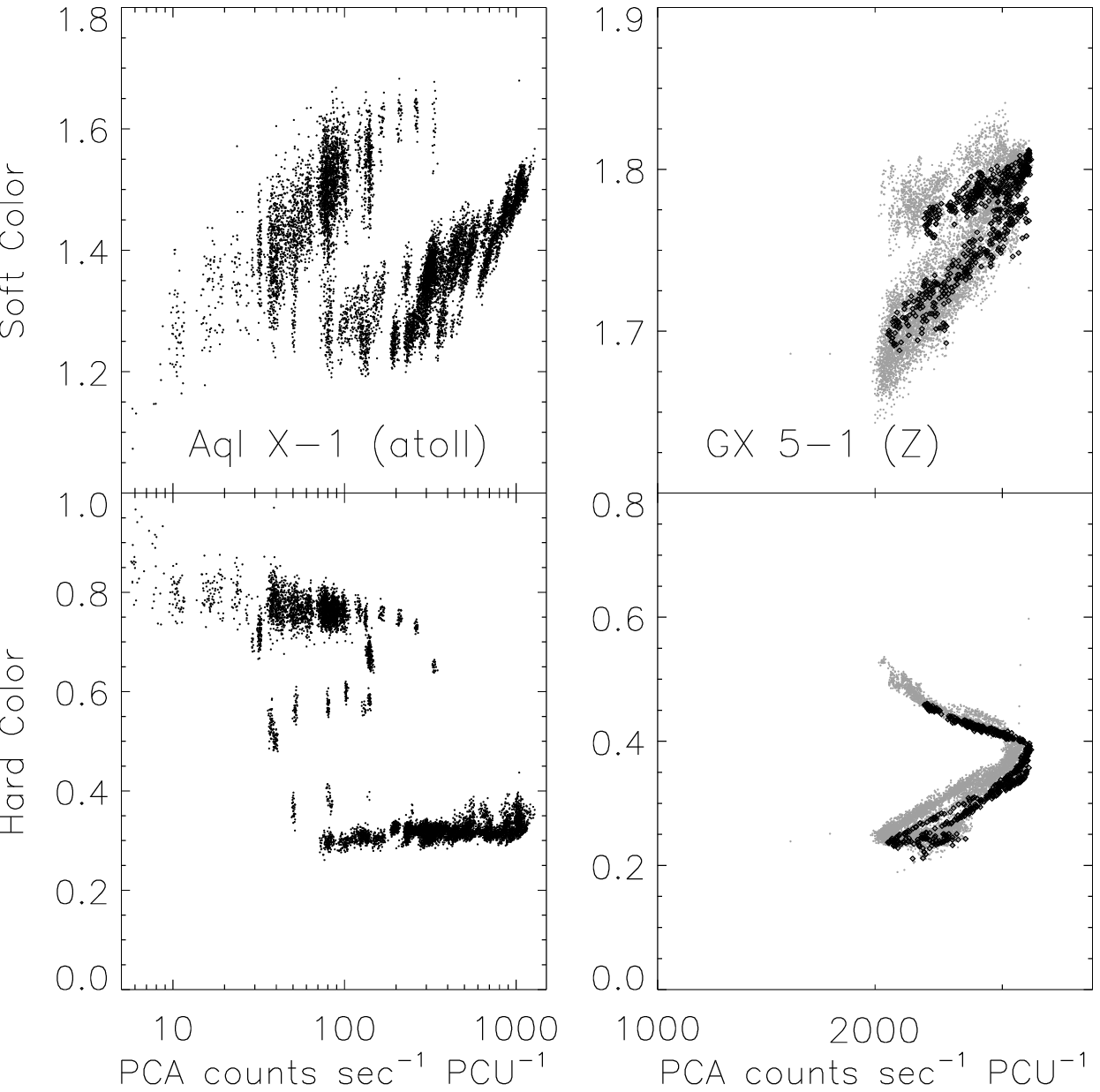,width=\linewidth}}
\caption{Plots of hard and soft color versus the PCA count rate between
2--18 keV for Aql~X-1 ({\it left}) and GX~5$-$1 ({\it right}). Notice that
narrower correlations between color and count rate are observed on time scales
of a day in Aql~X-1. For GX~5$-$1, we have highlighted observations spanning 20
days to account for global shifts in the position of the 
colors on longer time scales, as in Figure~1.}
\label{hid}
\end{inlinefigure}

\noindent
Z on short time scales in Figure~\ref{cconly}, 
in order to allow an easy comparison between the short and long-term spectral 
variability. As has been previously noted, the width of the track 
perpendicular to the direction of motion is much smaller in 
traditional Z 
sources than in atoll sources when observed on time scales of 
days \cite[e.g][]{vdk95}. On the 
five-year time scales of Figure~\ref{cconly}, both the
Z and atoll sources exhibit 10--20\% variations in the colors perpendicular 
to the direction of motion that traces the Z \citep[e.g.][]{kul94}. 

It is interesting to note 
that the color-color diagram of GX~13$+$1 in Figure~\ref{cconly}
resembles that of the Z source GX~17$+$2 more than 
those of the other atoll sources. Although GX~13+1 was classified an 
atoll source because it lacked strong QPOs \citep{hv89}, \rxte\
observations have revealed weak QPOs similar to those of Z sources
\citep{hom98}, and several observations have revealed similarities
between the X-ray \citep{sht89} and IR \citep{ban99} spectra of 
GX~13+1 and other Z sources. GX~13+1 appears to exhibit both Z and atoll 
properties, and may prove important for understanding what causes 
the distinctions between the two classes of source.

We next examine how these colors evolve versus intensity and,
for the atoll sources, versus time. In Figure~\ref{hid} we have plotted 
the hard 
and soft color 
as a function of the 2--18~keV PCA count rates for the atoll source 
Aql~X-1 and the Z source GX~5$-$1. In Figure~\ref{cctime} we have plotted
the X-ray colors and the 2-12~keV X-ray intensity from the \rxte\ ASM 
from the atoll source 4U~1705$-$440 over
a span of 200 days. It is well-known that Z sources trace their color-color
diagrams smoothly with time in less than a day \citep[e.g.][]{kul94,vdk95}. 
These data are the best-sampled examples from the LMXBs in 

\begin{inlinefigure}
\centerline{\epsfig{file=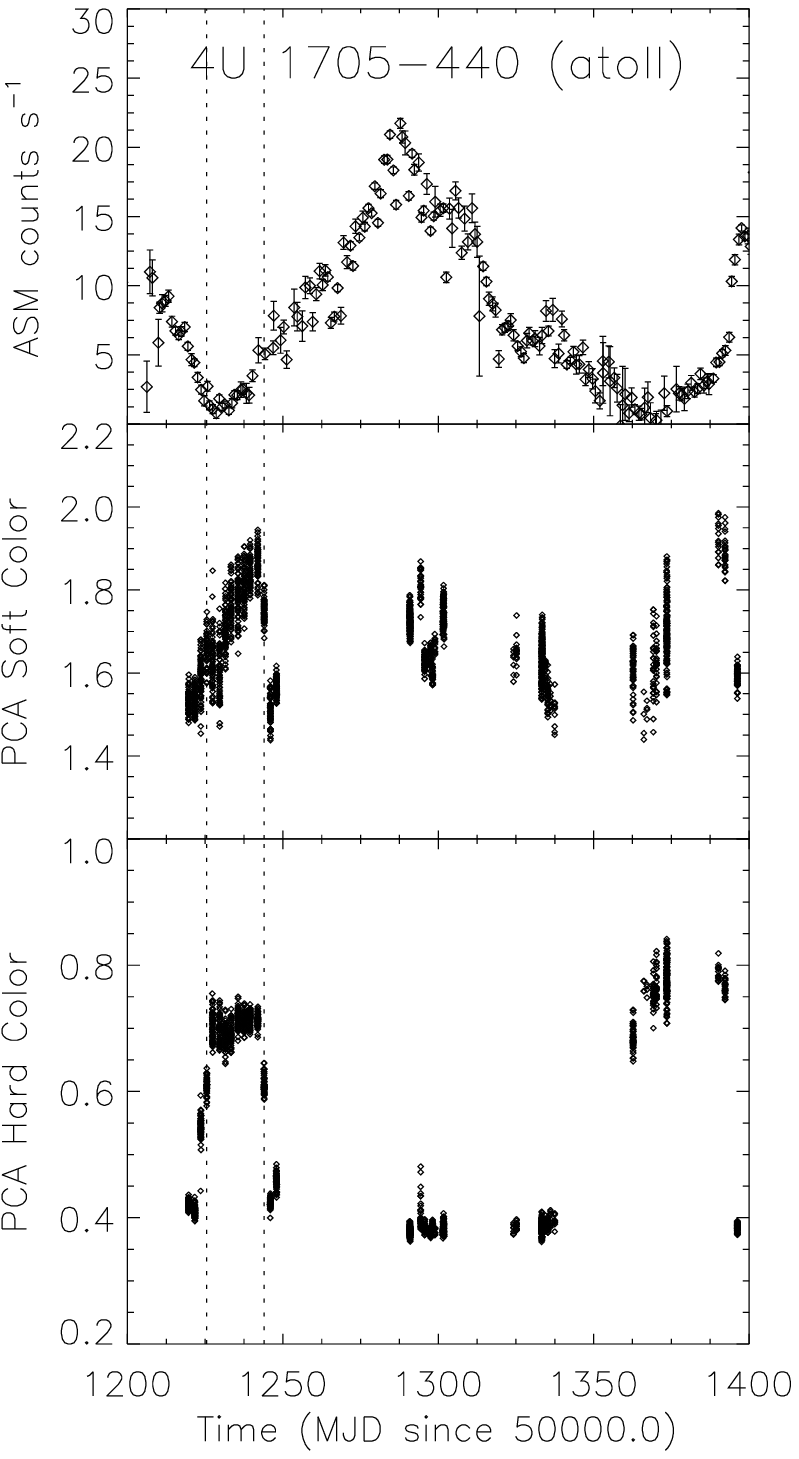,width=0.65\linewidth}}
\caption{The evolution of the X-ray intensity and colors as a function
of time for 4U~1705$-$440. {\it Top Panel:} X-ray intensity measured with 
the ASM (2--12 keV). {\it Middle and Bottom Panels:} Soft and hard X-ray
colors. The diagonal portion of the color-color diagram appears as the 
rapid change in hard color (see dashed lines), which occurs at a higher 
count rate in the rise of an outburst than in the decay.}
\label{cctime}
\end{inlinefigure}

\noindent
Table~\ref{stats}.

Figures~\ref{hid} and \ref{cctime} indicate that atoll sources also trace 
their
Z-shaped track smoothly. The horizontal portion at the top of the color-color
track of the atoll sources (with a hard color of about 0.8 in 
Figure~\ref{cconly}) is traced from left to right 
as the intensity 
increases by more than a factor of 10. For Aql~X-1 and 4U~1608-522, 
it is traced on time scales of days to months as the source rises from (or 
decays to) below the PCA detection threshold, with a factor of 250 change in
intensity. On the 
other hand, GS~1826$-$34 has remained in a similar faint and hard state 
for the six years of \rxte\ monitoring.

The diagonal branch is traced on time scales of days, and only has been sampled
well in time in a few instances (e.g. Figure~\ref{cctime}).
As an atoll source moves down along the diagonal portion of the color-color 
track, the count rate usually increases by a factor of 2 (and vice versa; 
see Figure~\ref{cctime}). However, during a recent 
unusual outburst of \aqlxone\ \citep{bai01} the count rate decreased by 
a factor of 2 while moving down along the diagonal branch (not shown).
The overall intensity on this branch is higher in the rise of an outburst 
(100--300 counts s$^{-1}$ PCU$^{-1}$ for Aql~X-1 in 
Figure~\ref{hid}) than in the decay 
($<$100 counts s$^{-1}$ PCU$^{-1}$ for Aql~X-1; see also Figure~\ref{cctime}). 
Since the soft color increases with intensity along the top branch 
(Figure~\ref{hid}), the value of soft color at which the diagonal and top
branches connect also could be smaller in the rise of an outburst than 
in the decay, as may be seen by comparing the soft colors near the dates 
indicated by dashed lines in Figure~\ref{cctime}.
Similar hysteresis has been observed from transient LMXBs containing black 
holes \citep{miy95}.

The bottom portion of the color-color track from the atoll sources
(hard color of 0.3)
is traced from left to right as the count rate increases once again by 
a factor of 10 (from 70--1500 counts 
s$^{-1}$ PCU$^{-1}$ in Figure~\ref{hid}). This branch is traced on time scales
of days to weeks in the transient sources, while Ser~X-1 
has remained in this soft state throughout the \rxte\ mission.

The color-color diagrams of Z sources are known to be traced smoothly, 
without jumping between branches, but the color changes are not accompanied 
by large variations in the X-ray intensity (right panels of Figure~\ref{hid}).
The top portion of the color-color track from GX~5$-$1
(hard color of 0.4 in Figure~\ref{cconly}) is traced as the count rate 
increases by 70\% (Figure~\ref{hid}). On the diagonal portion 
of the color-color track of
GX~5$-$1, the count rate returns to its lowest
values. In atoll sources,
the intensity is nearly constant. On the bottom of the color-color track, 
the count rate 
of GX~5$-$1 steadily increases yet again by 70\%. 
The intensity from other Z sources changes by up to a factor of 
three on this track \citep[e.g.,][for GX 17$+$2]{hom01}.

The hardness-intensity diagram of the atoll source 
Aql~X-1 in Figure~\ref{hid} also reveals important 
sub-structure on short time scales that is common to all of the 
atoll sources in Table~\ref{stats}. This structure is most
evident in the plot of soft color against count rate, where narrow, 
parallel tracks are traced by observations within single 
days. At
low intensities (less than 200 counts s$^{-1}$ 
PCU$^{-1}$), the soft 
color varies as the count rate remains nearly constant, while at the 
highest intensities, the soft color and count rate 
both increase significantly together. On the color-color diagram in 
Figure~\ref{cconly}, these daily tracks are generally aligned perpendicular to 
motion along the Z, and thus broaden the color-color tracks of the atoll 
sources. In contrast, the parallel tracks from the Z sources are formed
when the entire Z shifts \citep{vdk00}.

\section{Discussion} 

We have demonstrated that the color-color diagrams of both Z and atoll
sources have similar three-branched shapes 
(Figures~\ref{cconly}), but 
that the range of X-ray intensity and the time scale over which 
the diagrams are traced are one to two orders of magnitude larger in the atoll 
sources (Figures~\ref{hid} and \ref{cctime}). We note that similar conclusions
have been reported independently by \citet{gd02}. There are also
significant spectral differences between the two classes of 
source \citep{sht89,cs97,bar00}. The spectra of Z sources are always soft, and 
can be described by the sum of a cool (1~keV) black 
body and Comptonized emission from warm (5~keV) optically thick electrons
\citep[e.g.,][]{cs97, dis00}. Spectral changes along the Z are 
quite subtle \citep[e.g.,][]{sht89}. In contrast, changes in the energy 
spectra of atoll sources are dramatic. While soft energy spectra
resembling those of Z sources are characteristic of atoll sources at 
the bottom of their color-color diagrams \citep[e.g.,][for Ser~X-1]{oos01}, 
a hard state with a $\Gamma \approx 1.8$ power-law energy spectrum 
between 2--100~keV occurs at the top of this diagram when the sources are 
faint \citep[e.g.,][for GS~1826$-$238]{bar00}. Such a hard spectrum is not 
observed from Z sources, probably because they are not observed at
low luminosities (Table~\ref{stats} and Figure~\ref{hid}).

Recent work has suggested that the timing properties of 
Z and atoll sources also may have interesting similarities.
It has long been suggested that the broad-band noise in power spectra of 
Z and atoll sources can probably be described by similar components 
\citep{vdk95}, and detailed studies of \rxte\ data have shown that 
the frequencies of QPOs and band-limited noise exhibit similar correlations
in both types of sources \citep{wk99,pbk99}. However, whether the 
timing properties are correlated with the branches
of the color-color diagrams in a similar manner in each class of source
remains to be investigated.

The timing properties of neutron star LMXBs eventually may allow us to 
understand
what drives the spectral variability. Several studies have found
that the kHz QPO frequencies are well-correlated with position on 
the color-color diagram in both Z \citep[e.g.,][]{hom01} and atoll 
\citep[e.g.,][]{vst01b} sources. 
Examining Figure~\ref{hid}, the short-term correlations 
between the colors and PCA count rate are extremely similar to the ``parallel 
tracks'' observed in comparisons of the frequencies of kilohertz 
quasi-periodic oscillations (kHz QPO) with the PCA count rate 
\citep{vdk00,vdk01}. 
The parallel tracks are observed from kHz QPOs both in studies of 
individual sources \citep{men99}, and in comparing the frequencies of QPOs 
from Z and atoll 
sources that span a wide range of luminosity \citep{for00}.
Based upon these parallel tracks, \citet{vdk01} has suggested that 
the kHz QPO frequencies are determined by a feedback process that is sensitive
to deviations in the accretion rate about its value averaged over a few days. 
Further work is needed to determine whether 
the same feedback mechanism can operate to produce spectral variations 
in both Z and atoll sources, on widely different time scales and over ranges 
in luminosity that differ by factors of 100.

\acknowledgments
We are grateful for questions and comments from Michiel van der Klis
and Erik Kuulkers that helped in clarifying these results.
This work was supported by NASA, under contract NAS 5-30612 and grant 
NAG 5-9184, and has made use of data obtained from the High Energy 
Astrophysics Science Archive Research Center (HEASARC), 
provided by NASA's Goddard Space Flight Center

\begin{deluxetable}{llcccc}
\tabletypesize{\small}
\tablecolumns{6}
\tablewidth{0pc}
\tablecaption{Basic Properties of Neutron Star LMXBs\label{stats}}
\tablehead{
\colhead{Source} & \colhead{Type} & \colhead{$L_{\rm X}$\tablenotemark{b}} & 
\colhead{$D$\tablenotemark{c}} & \colhead{$N_{\rm H}$\tablenotemark{d}} & 
\colhead{Variability\tablenotemark{e}}
}
\startdata
4U~1608$-$522 & A,b & 0.1 & 4 & 1.5 & 2000\\ 
Aql~X-1 & A,b & 0.1 & 3 & 0.5 & 1000\\ 
4U~1705$-$440 & A,b & 6 & 11 & 1.1 & 80\\ 
4U~1728$-$34 & A,b & 0.4 & 4 & 1.7 & 10\\
Ser~X-1 & A,b & 4 & 8\tablenotemark{f} & 0.5 & 5\\ 
4U~1820$-$303 & A,b & 5 & 7.5 & 2.6 & 4\\ 
GX~13$+$1 & A,b & 5 & 7\tablenotemark{f} & 2.5 & 4\\ 
GS~1826$-$238 & A?,b & 0.3 & 6\tablenotemark{g} & 0.5\tablenotemark{g} & 2\\
Cyg~X-2 & Z,b & 16 & 12 & 0.3 & 6\\ 
GX~17$+$2 & Z,b & 12 & 8 & 1.7 & 3\\ 
GX~5$-$1 & Z & 18 & 7 & 2.5 & 3\\ 
GX~340$+$0 & Z & 16 & 10 & 5.0 & 3\\ 
\enddata
\tablenotetext{a}{Previous classifications of each source. A --- atoll source,
Z --- Z source, b --- burster.}
\tablenotetext{b}{Mean unabsorbed 2-12~keV luminosity in 
units of 10$^{37}$ erg s$^{-1}$, inferred from the ASM assuming a 
$\Gamma = 2.5$ power law spectrum.}
\tablenotetext{c}{Distance in kpc, taken from Ford et al. (2000) except 
where noted}.
\tablenotetext{d}{Absorption column in units of $10^{22}$ cm$^{-2}$, taken 
from Christian \& Swank (1995) except where noted.}
\tablenotetext{e}{$F_{\rm max}/F_{\rm min}$, 2--18 keV from the PCA.} 
\tablenotetext{f}{From Christian \& Swank (1995).}
\tablenotetext{g}{From in 't Zand et al. (1999).}
\end{deluxetable}

\end{document}